\documentclass[fleqn,10pt]{wlscirep}
\usepackage[utf8]{inputenc}
\usepackage[T1]{fontenc}
\usepackage{graphicx}% Include figure files
\usepackage{dcolumn}% Align table columns on decimal point
\usepackage{bm}% bold math
\usepackage{color}

\title{Controls of a superconducting quantum parametron under a strong pump field}

\author[1,*]{Shumpei Masuda}
\author[1]{Toyofumi Ishikawa}
\author[1]{Yuichiro Matsuzaki}
\author[1]{Shiro Kawabata}
\affil[1]{Research Center for Emerging Computing Technologies (RCECT), National Institute of Advanced Industrial Science and Technology (AIST), 1-1-1, Umezono, Tsukuba, Ibaraki 305-8568, Japan}
\affil[*]{shumpei.masuda@aist.go.jp}

\begin{abstract}
Pumped at approximately twice the natural frequency, a Josephson parametric oscillator called parametron or Kerr parametric oscillator shows self-oscillation. 
Quantum annealing and universal quantum computation using self-oscillating parametrons as qubits were proposed.
However, controls of parametrons under the pump field are degraded by unwanted rapidly oscillating terms in the Hamiltonian, which we call non-resonant rapidly oscillating terms (NROTs) coming from the violation of the rotating
wave approximation.
Therefore, the pump field can be an intrinsic origin of the imperfection of controls of parametrons.
Here, we theoretically study the influence of the NROTs on the accuracy of controls of a parametron: a cat-state creation and a single-qubit gate.
It is shown that there is a trade-off relationship between the suppression of the nonadiabatic transitions and the validity of the rotating wave approximation in a conventional approach. We also show that the tailored time dependence of the detuning of the pump field can suppress both of the nonadiabatic transitions and the disturbance of the state of the parametron due to the NROTs.
\end{abstract}
\begin{document}

\flushbottom
\maketitle
\thispagestyle{empty}

%\noindent Please note: Abbreviations should be introduced at the first mention in the main text \UTF{2013} no abbreviations lists. Suggested structure of main text (not enforced) is provided below.

\section*{Introduction}
Parametric phase-locked oscillators \cite{Onyshkevych1959}, which are also called parametrons \cite{Goto1959}, can store binary digital information as the phase of the self-oscillation when they are driven via a periodic modulation of their circuit element.
Parametrons were actually operated as classical bits in digital computers in 1950s and 1960s until the transistor acquired the solid stability.
More recently, parametrons were revived in the nanoelectromechanical, optical and the superconducting circuit systems.
Basic bit operations have been demonstrated in a nanoelectromechanical system using a electromechanical resonator~\cite{Mahboob2008}, and the Ising machine based on optical parametron has been proposed~\cite{Wang2013}. 
To see the quantum nature of the parametron, the nonlinearity should be sufficiently large compared to the decay rate. 
The nonlinearity smaller than the decay rate gives rise to the appearance of classical dynamics of the system~\cite{Wilson2010}.
The quantum regime with the nonlinearity larger than the decay rate has been studied theoretically~\cite{Kinsler1991,Wustmann2013,Zhang2017} and experimentally  
\cite{Ding2017,Wang2019}. We consider this quantum regime in this paper.

The parametron was applied to the qubit readout~\cite{Yamamoto2014,Yamamoto2016} in circuit QED architectures which are promising platform of quantum information processing \cite{Nakamura1999,Krantz2019,Blais2020}.
Quantum annealing \cite{Goto2016,Nigg2017,Puri2017}
and universal quantum computation \cite{Goto2016b}, which utilize the quantum nature of parametrons in a superconducting circuit, have been proposed.
Recently, the bias-preserving gates~\cite{Puri2020} and single-qubit operations~\cite{Grimm2020} were studied theoretically and experimentally. 
Exponential increase of the bit-flip time with the cat size was also observed \cite{Lescanne2020}.

Under the pump field oscillating at approximately twice its natural frequency, a superconducting quantum parametron (we refer parametron hereafter) can work as a qubit in contrast to transmons and flux qubits which do not require an oscillating pump field to realize an effective two-level system.

The decay from the parametron causes the decoherence of the qubit states~\cite{Puri2017b}.
In order to avoid the decoherence, we need controls much faster than the decay rate. 
For such rapid controls, we require a large pump field to avoid unwanted nonadiabatic transitions~\cite{Goto2016b}. However, the strong pump field can be an origin of the degradation of qubit operations. 
Such a trade-off relationship has been overlooked in earlier studies on the parametron.

In this paper, we study the effect of the strong pump field to the operations of a parametron in the quantum regime assuming that the operation time is much shorter than the coherence time.
First, in order to quantitatively assess the feasibility of superconducting parametron for quantum applications, we study the effect of the unwanted non-resonant rapidly oscillating terms (NROTs) in the Hamiltonian on the accuracy of the creation of a cat state. 
%and a single-qubit gate along the $x$ axis.
It is shown that there is a trade-off relationship between the suppression of the nonadiabatic transitions and the validity of the rotating wave approximation in a conventional approach \cite{Goto2016,Goto2016b}. Second, we also show that the tailored time dependence of the detuning of the pump field can suppress both the nonadiabatic transitions and the disturbance of the state of a parametron due to the NROTs. 
Finally, we study the effect of the NROTs on an $R_x$ gate.

\section*{Model}
\label{Model}
\begin{figure}
\begin{center}
\includegraphics[width=10cm]{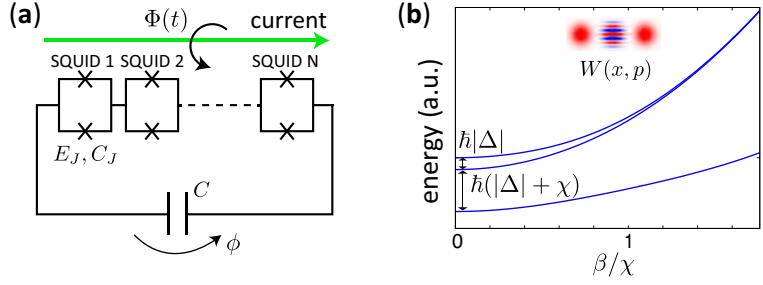}
\end{center}
\caption{
(a) Schematic of a superconducting quantum parametron. $E_J$ and $C_J$ are the Josephson energy of a single SQUID and the capacitance of a single Josephson junction, respectively. $C$ denotes the capacitor shunting the SQUID array. $\phi$ is the overall phase across the junction array.  $\Phi(t)$ is the external magnetic flux threading the SQUIDs. 
(b) Energy level diagram as a function of pump strength for $\Delta<0$ and $\chi>0$.
Because $\Delta<0$, the highest energy level is the vacuum state for $\beta=0$.
The inset is a typical image of the Wigner function of the highest energy level for large $\sqrt{2\beta/\chi}(\simeq 2.5)$.
The top two curves overlap as $\beta$ is sufficiently large.}
\label{KPO_system_6_1_20}
\end{figure}

We consider a parametron composed of a SQUID-array resonator with $N$ SQUIDs (Fig.~\ref{KPO_system_6_1_20}(a)) which was implemented in Ref.~\citenum{Wang2019}.
The effective Hamiltonian of the system is represented as~\cite{Wang2019} 
\begin{eqnarray}
H= 4E_C n^2 - NE_J[\Phi(t)] \cos\frac{\phi}{N},
\label{H_KPO_4_16_20}
\end{eqnarray}
where $\phi$ and $n$ are the overall phase across the junction array and its conjugate variable, respectively.
$E_J$ is the Josephson energy of a single SQUID.
The effective Hamiltonian with a single degree of freedom, $\phi$, is valid when the Josephson energy $E_J$ is much greater than the charging energy of a single junction~\cite{Frattini2017}.
$E_C$ is the resonator's charging energy including the contributions of the junction capacitance $C_J$ and the shunt capacitance $C$, and can be extracted from measurements and also can be calculated with finite-element capacitance simulations~\cite{Wang2019}. 
The Josephson energy is periodically modulated by the external magnetic flux, $\Phi(t)$, threading the SQUIDs as
$E_J(t)=E_J+\delta E_J \cos\omega_p t$.

Taking into account up to the 4th order of $\phi/N$ in Eq.~(\ref{H_KPO_4_16_20}), we obtain an approximate Hamiltonian
\begin{eqnarray}
\frac{H}{\hbar} &=& \omega_c^{(0)} \Big{(} a^\dagger a + \frac{1}{2} \Big{)}
- \frac{\chi}{12} (a + a^\dagger)^4
 + \Big{[} - \frac{N\delta E_J}{\hbar}  +
 2\beta (a + a^\dagger)^2  - \frac{2\chi \beta}{3\omega_c^{(0)}} (a + a^\dagger)^4 \Big{]}
\cos\omega_p t.
\label{H_9_1_20}
\end{eqnarray}
where $\omega_c^{(0)} = \frac{1}{\hbar}\sqrt{8E_CE_J/N}$, $\chi=E_C/\hbar N^2$ and $\beta = \omega_c^{(0)} \delta E_J/ 8E_J$. 
Here, $\beta$ corresponds to the pump strength.
The annihilation operator $a$ is related to $n$ and $\phi$ as
$n = -in_0(a-a^\dagger)$ and $\phi = \phi_0 (a+a^\dagger)$
with $n_0^2=\sqrt{E_J/32 N E_C}$ and $\phi_0^2 = \sqrt{2NE_C/E_J}$.
For the expansion of Eq.~(\ref{H_KPO_4_16_20}), we considered the parameter regime, where $\phi_0/N = 2\sqrt{\chi/\omega_c^{(0)}}$ is sufficiently smaller than unity so that the approximation is valid.
We took into account up to the forth order of $\phi/N$ to see the effect of the Kerr nonlinearity, which is important for a parametron.
We neglect the last term in Eq.~(\ref{H_9_1_20}) assuming $\chi\beta\ll  \omega_c^{(0)}$, and drop c-valued terms to obtain
\begin{eqnarray}
\frac{H}{\hbar} = \omega_c^{(0)} a^\dagger a 
- \frac{\chi}{12} (a + a^\dagger)^4
+ 2\beta (a + a^\dagger)^2
\cos\omega_p t.
\end{eqnarray}
Moving into a rotating frame at the frequency of
$\omega_p/2$, the Hamiltonian is written as
\begin{eqnarray}
\frac{H}{\hbar} &=& \Big{(}\omega_c^{(0)}-\omega_p/2 \Big{)} a^\dagger  a 
- \frac{\chi}{12} (a e^{-i\frac{\omega_p}{2}t} + a^\dagger e^{i\frac{\omega_p}{2}t} )^4 + 2\beta (a e^{-i\frac{\omega_p}{2}t}  + a^\dagger e^{i\frac{\omega_p}{2}t})^2
\cos\omega_p t.
\label{Hrf_9_1_20}
\end{eqnarray}
When we neglect all the oscillating terms such as $a^2e^{-2i\omega_p t}$ which are called NROTs,
we obtain an approximate Hamiltonian (rotating wave approximation),
\begin{eqnarray}
\frac{H_{\rm RWA}} {\hbar} = \Delta a^\dagger a - \frac{\chi}{2} a^\dagger a^\dagger a a
+\beta(a^2 + a^{\dagger 2}),
\label{H_6_8_20}
\end{eqnarray}
where $\Delta = \omega_c^{(0)}-\chi -\omega_p/2$. 
We compare the results for the Hamiltonians in Eqs.~(\ref{Hrf_9_1_20}) and~(\ref{H_6_8_20}) in the following sections.
We neglect the decay and the dephasing to highlight the effect of the NROTs assuming that the decay and the dephasing time is sufficiently longer than the duration of the controls.

Figure~\ref{KPO_system_6_1_20}(b) shows a schematic of the energy level diagram of the Hamiltonian~(\ref{H_6_8_20}).
The vacuum state is the highest energy level in the rotating frame when $\beta=0$. 
The highest and the second highest energy levels for sufficiently large $\beta/\chi$ are represented as
\begin{eqnarray}
|\varphi_0\rangle &\simeq& \frac{|-\alpha\rangle + |\alpha\rangle}{\sqrt{2}},\nonumber\\ 
|\varphi_{1}\rangle &\simeq& \frac{|-\alpha\rangle - |\alpha\rangle}{\sqrt{2}},
\label{cat_6_23_20}
\end{eqnarray}
respectively, with coherent states, $|-\alpha\rangle$ and $|\alpha\rangle$, where $\alpha = \sqrt{(2\beta+\Delta)/\chi}$~\cite{Goto2019a}, and $|\Delta|$ is much smaller than $\beta$.
These coherent states, $|-\alpha\rangle$ and $|\alpha\rangle$, can be used as a qubit for quantum annealing and universal quantum computation \cite{Goto2016,Goto2016b}. 
Thus, the creation of predetermined states such as cat states in Eq.~(\ref{cat_6_23_20}) is of importance for quantum information processing.

In this paper, we consider the case that $\Delta\le 0$. If $\Delta$ is positive, the vacuum state is not the highest energy level in the rotating frame when $\beta=0$, and the vacuum state is driven to a state different from 
$|\varphi_0\rangle$ as the pump field is ramped~\cite{Zhang2017}.

\section*{Results}
\label{Results}
We examine the effect of the NROTs on the creation of a cat state, $|\varphi_0\rangle$, 
and on an accuracy of a single-qubit gate along the $x$ axis ($R_x$ gate).
We solve the time-dependent Schr\"{o}dinger equation with a fourth-order Runge-Kutta integrator with the
time step of 0.025~fs in the following numerical simulations. 

\subsection*{Creation of a cat state}
\label{Creation of a cat state}
We assume that the system is in the vacuum state and $\beta=0$ at $t=0$; and
$\beta$ is gradually increased for $0\le t\le T$.
The quantum adiabatic theorem states that the system remains in the highest energy level if 
$\beta$ is increased slowly enough. 
Thus, the population of the highest energy level, $p_0$, is unity if the evolution is completely adiabatic.
We set the time dependence of $\beta$ as
\begin{eqnarray}
\beta(t) =
\left\{ 
\begin{array}{cc}
\beta_0 t / T  \ & {\rm for} \ \ 0\le t \le T,\\
\beta_0  \ & {\rm for}  \  \ t > T.
\end{array}
\right.
\label{beta_9_3_20}
\end{eqnarray}
(We consider a linear ramp of $\beta$ for simplicity.)
We define the fidelity of the control as $p_0(t)$ for $t>T$.

\begin{figure}
\begin{center}
\includegraphics[width=10cm]{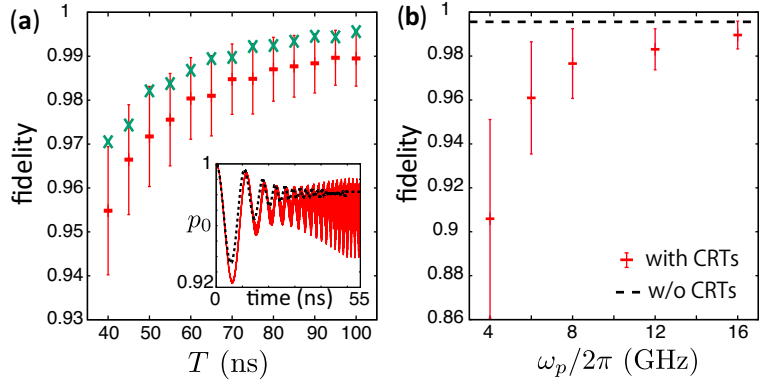}
\end{center}
\caption{
(a) Fidelity of the creation of a cat state as a function of $T$ for the dynamics with (red) and without (green) the NROTs, where the error bars represent the standard deviation calculated using the data for $t>T$.
The used parameters are $\beta_0/2\pi = 200$~MHz, $\Delta/2\pi=-6.7$~MHz, $\omega_p/2\pi=16$~GHz and $\chi/2\pi=68$~MHz. 
The inset shows the time evolution of $p_0$ for $T=50$~ns.
The red solid and the black dashed curves are for with and without the NROTs, respectively.
We chose $\Delta$ to be the same as the previous study~\cite{Wang2019}.
%\textcolor{blue}{When $\Delta$ is positive, $|n(\ne0)\rangle$ becomes the highest level for $\beta=0$ in contrast to our case.}
(b) Fidelity as a function of $\omega_p/2\pi$ for $T=100$~ns. 
We use the same value for $\beta_0$, $\chi$ and $\Delta$ as panel (a), while $\omega_c^{(0)}$ is changed so that $\Delta$ is unchanged (Note that $\omega_c^{(0)} = \omega_p/2 + \Delta + \chi$).
The dashed line corresponds to the dynamics without the NROTs.
}
\label{pop_data_6_8_20}
\end{figure}
Figure~\ref{pop_data_6_8_20}(a) shows the fidelity of the control as a function of $T$.
The fidelity for short $T$ is lowered due to unwanted nonadiabatic transitions in the dynamics without NROTs.
In the dynamics with the NROTs, the fidelity is even lower and keeps fluctuating after the ramp of the pump field.
The standard deviation of the fluctuation of $p_0$ for $t>T$ is considerably large even for $T=$ 100~ns where the nonadiabatic transitions are negligible.
The fluctuation becomes large when $T$ is short because of the large population of the lower levels.
Figure~\ref{pop_data_6_8_20}(b) shows the fidelity as a function of $\omega_p$. In this numerical simulation, $\omega_c^{(0)}$ is changed with $\omega_p$ so that the detuning is fixed. 
It is seen that, as $\omega_p$ increases, the fidelity is increased and the fluctuation of $p_0$ is suppressed.  
This comes from the fact that the rotating wave approximation becomes more accurate as we increase $\omega_p$ and  $\omega_c^{(0)}$.

\begin{figure}
\begin{center}
\includegraphics[width=10cm]{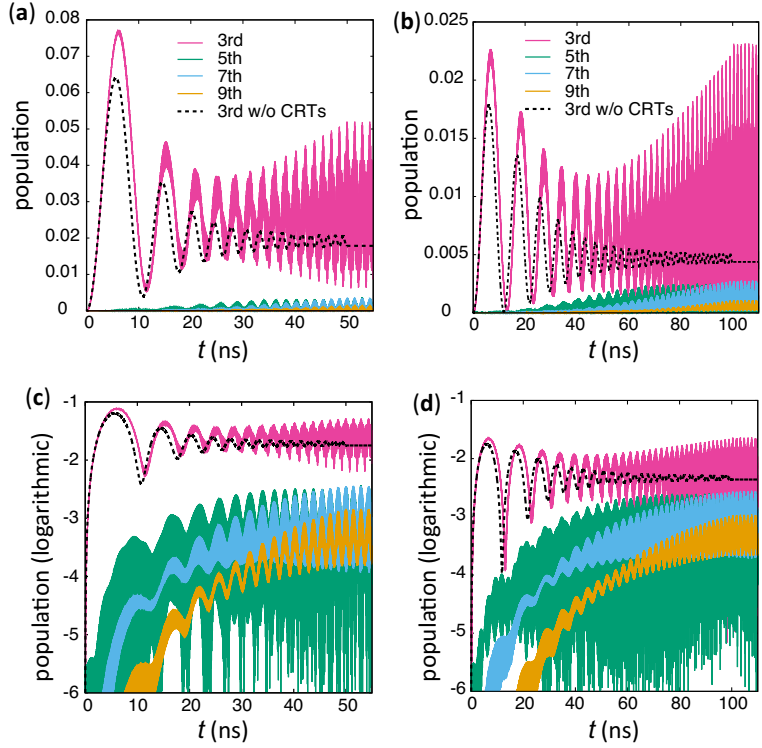}
\end{center}
\caption{
Time dependence of the population of the third, fifth, seventh and the nineth highest levels during the cat-state creation for $T=50$~ns (a) and 100~ns (b).
The population of the second, fourth, sixth, $\cdots$ levels is vanishing because of the difference of the parity.
The dotted curves represent the population of the third highest level in the dynamics without NROTs.
The used parameters are the same as Fig.~\ref{pop_data_6_8_20}(a).
Panels (c) and (d) are the same things as panels (a) and (b), respectively, but with the vertical axis in the logarithmic scale.
}
\label{pop_com_9_4_20_3}
\end{figure}

The time dependences of the population of lower levels are shown for $T=50$~ns and 100~ns in Figs.~\ref{pop_com_9_4_20_3}(a,c) and \ref{pop_com_9_4_20_3}(b,d) respectively.
In the case without the NROTs, the third highest level is populated due to the nonadiabatic transition while the population of the other lower levels are approximately zero ($e.g.$, the population of the fifth highest level is less than $10^{-5}$ and $10^{-6}$ at $t=T$ for the parameters used in Fig.~\ref{pop_com_9_4_20_3}(a,c) and Fig.~\ref{pop_com_9_4_20_3}(b,d), respectively).
The population of the second, fourth, sixth, $\cdots$ levels is vanishing because of the parity difference from the highest level.
On the other hand, the other lower levels with the same parity as the highest level are also populated in the dynamics with the NROTs as apparently seen in Fig.~\ref{pop_com_9_4_20_3}.
The fluctuating population of the third highest level is higher than that without the NROTs for the both values of $T$.
The oscillation of the populations saturates for $t>T$, when $\beta$ is constant.

We discuss the significance of our results here.
It is worth mentioning that we need a condition of $\langle -\alpha |\alpha \rangle \simeq 0$ to use the parametron as a qubit, and so $\beta/\chi$ should be sufficiently large. (The overlap, $\langle -\alpha |\alpha \rangle$, becomes negligible when $\beta/\chi$ is sufficiently large because $\langle -\alpha | \alpha \rangle = \exp[-2|\alpha|^2]$~\cite{Goto2016} and $\alpha=\sqrt{(2\beta+\Delta)/\chi}$.)
For this purpose, we could decrease $\chi $, but this leads us to a smaller energy gap between the eigenenergies of the Hamiltonian, which could induce more nonadiabatic transitions. 
An alternative approach to satisfy $\langle -\alpha |\alpha \rangle \simeq 0$ while supressing the nonadiabatic transitions could be a increase of $\beta $. 
However, as we showed in this subsection, a large $\beta $ could be another source of error due to the violation of the rotating wave approximation. 
Therefore, in the conventional approach, there is a trade-off relationship between the suppression of the nonadiabatic transitions and the validity of the rotating wave approximation, which was often overlooked in earlier works.

\subsubsection*{Suppression of nonadiabatic transitions}
\label{Suppression of nonadiabatic transitions}
In order to overcome the trade-off relationship discussed in the previous subsection,
we examine a way to enhance the fidelity of the creation of a cat state based on the time-dependent detuning~\cite{Goto2019P}.
We show that the fluctuation of the population of the target state due to the NROTs and the nonadiabitc transitions are greatly suppressed without increasing $\beta$ nor decreasing $\chi$.

In this method, we set the initial detuning large and decrease it to zero as
\begin{eqnarray}
\Delta(t) =
\left\{ 
\begin{array}{clc}
\Delta_0 (1- t / T)  \ & {\rm for} \ \ 0\le t \le T,\\
0  \ & {\rm for}  \  \ t > T.
\end{array}
\right.
\label{Delta_9_7_20}
\end{eqnarray}
The pump is ramped following Eq.~(\ref{beta_9_3_20}).
We set the initial detuning $\Delta_0/2\pi=-67$~MHz. 
The time-dependent detuning can be implemented by controlling $\omega_c^{(0)}$ depending on $E_J$ which can be controlled with the magnetic flux. Unwanted resultant change in $\beta$ can be compensated by changing $\delta E_J$. Alternatively, the time-dependent frequency of the pump field can be used for the implementation of the time-dependent detuning.

\begin{figure}
\begin{center}
\includegraphics[width=10cm]{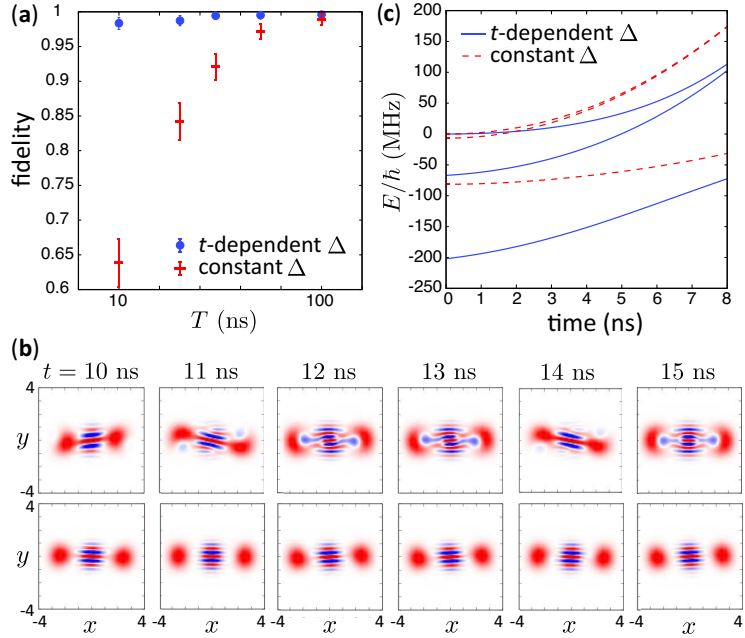}
\end{center}
\caption{
(a) Fidelity of the creation of a cat state as a function of $T$ with the time-dependent detuning in Eq.~(\ref{Delta_9_7_20}) (blue circles) and the constant detuning of $\Delta/2\pi=-6.7$~MHz (red bars). 
The NROTs are taken into account in the both dynamics.
The error bars represent the standard deviation which is calculated using the data for $t>T$. 
The used parameters are $\Delta_0/2\pi=-67$~MHz, $\beta_0/2\pi = 200$~MHz, $\omega_p/2\pi=16$~GHz and $\chi/2\pi=68$~MHz.
(b) Wigner functions for $t\ge T (=10~{\rm ns})$ in the controls with the constant (upper figures) and the time-dependent (lower figures) detunings.
The other parameters are the same as panel (a).
(c) The three highest eigenenergies of instantaneous $H_{\rm RWA}$ in Eq.~(\ref{H_6_8_20}) for the constant and the time-dependent detuning for $T=20$ ns. The other parameters are the same as panel (a).
}
\label{eng_cat_pop_9_3_20}
\end{figure}

Figure~\ref{eng_cat_pop_9_3_20}(a) represents the fidelity of the creation of a cat state as a function of $T$.
The modified method gives the fidelity considerably higher than the one with the constant detuning.
We have obtained the fidelity of more than 0.995 with the modified method for $T=50$~ns while the average fidelity for the control with the constant detuning is approximately 0.97.
We emphasize that the fluctuation of the fidelity is suppressed in the modified method as seen in the error bars of Fig.~\ref{eng_cat_pop_9_3_20}(a). 
We attribute this to the fact that the population of the lower levels are much smaller than the case with the constant detuning. Note that the NROTs, which couples the highest level to the other levels, weakly influence to the population of the highest level, when the population of the other levels are small.
Figure~\ref{eng_cat_pop_9_3_20}(b) represents the Wigner function \cite{Goto2016} for $t\ge T (=10~{\rm ns})$ in the controls with the constant and the time-dependent detunings.
The Wigner function is disturbed and time dependent in the control with the constant detuning for $t\ge T$, while in the modified method it is approximately stationary and coincides with that of the highest energy level of $H_{\rm RWA}$.
The results for the controls with different values of $\Delta_0$ are shown in Supplementary Section S1.

Figure~\ref{eng_cat_pop_9_3_20}(c) shows the three highest eigenenergies of instantaneous $H_{\rm RWA}$ in Eq.~(\ref{H_6_8_20}) for the constant and the time-dependent detuning.
The reader may consider that the nonadiabatic transitions occur when $t$ is large because the interval between the highest and the second highest levels become small. However, such transition does not occur because of the parity difference.
The major population transfer is from the highest level to the third highest level.

The enhancement of the fidelity in the modified method is explained as follows.
It is known that the adiabatic condition:
\begin{eqnarray}
h_{mn}(t) = \hbar|\langle \varphi_n(t) | \dot\varphi_m(t) \rangle| / |E_n(t)-E_m(t)| \ll 1
\label{hmn_12_15_20}
\end{eqnarray}
should be satisfied to suppress the nonadiabatic transition between levels $m$ and $n$, where $E_m$ is an eigenvalue of the instantaneous $H_{\rm RWA}$, and $m\ne n$. 
The state of the highest level of $H_{\rm RWA}$ changes drastically from the zero photon state to a superposition of Fock states as the pump is ramped in the small pump regime. 
The introduced large detuning in the small pump regime makes slow the rate of the change of the highest level, and makes the denominator of Eq.~(\ref{hmn_12_15_20}) large. Thus, the dynamics is well approximated by the adiabatic dynamics (nonadiabatic transitions are suppressed).
On the other hand, the rate of the change of the highest level is slow for the large pump regime compared to the small pump regime.
Therefore, the detuning can be gradually turned off.

\begin{figure}[h]
\begin{center}
\includegraphics[width=6.5cm]{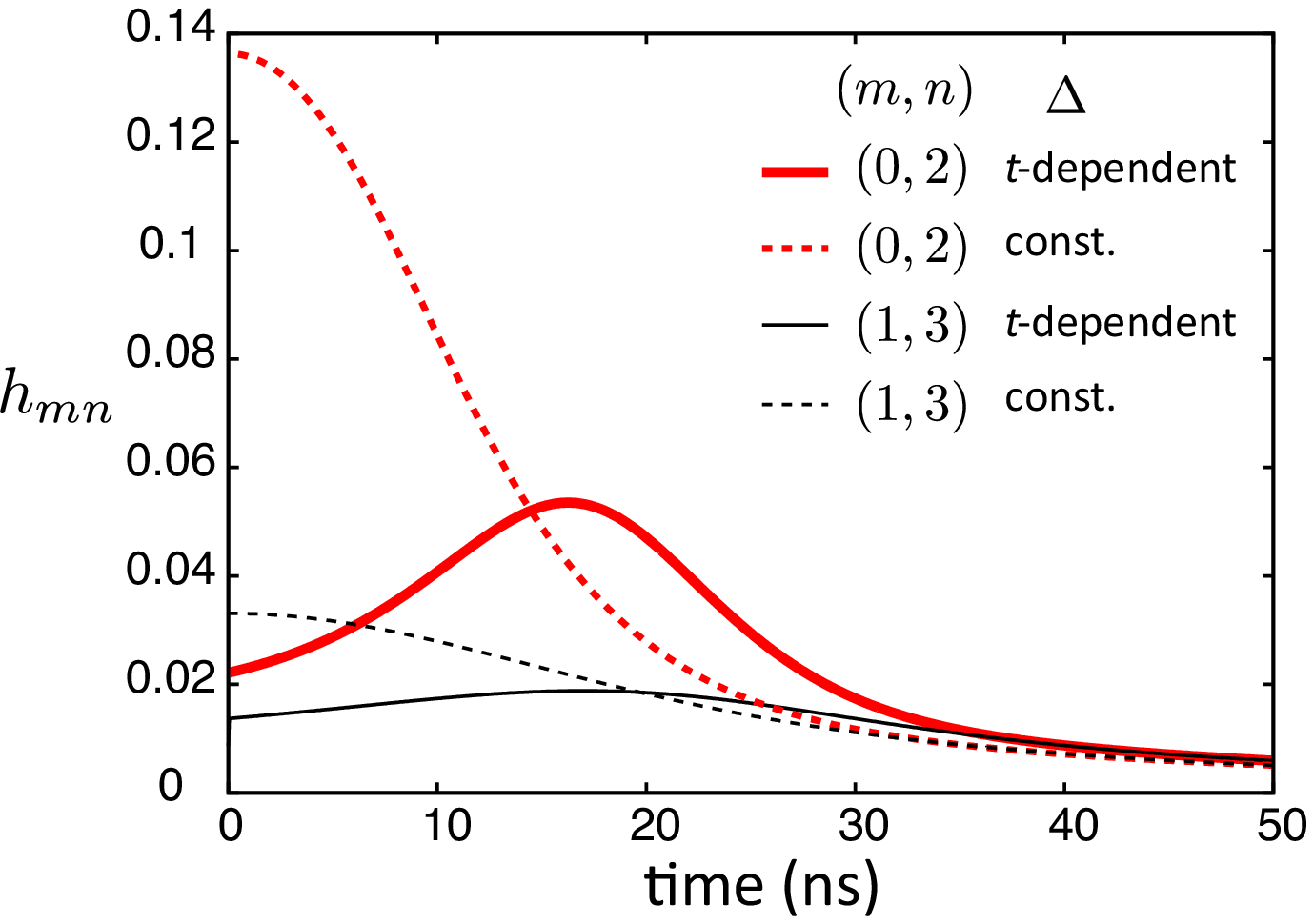}
\end{center}
\caption{
Time dependence of $h_{mn}$ for the creation of a cat state for $T=50$~ns with the time-dependent detuning in Eq.~(\ref{Delta_9_7_20}) and the constant detuning.
Other parameters are the same as Fig.~\ref{pop_data_6_8_20}(a).
}
\label{Hcd_com_9_7_20}
\end{figure}

Figure~\ref{Hcd_com_9_7_20} shows the time dependence of $h_{mn}$ during the creation of a cat state with the time-dependent detuning in Eq.~(\ref{Delta_9_7_20}) and the constant detuning.
It is seen that $h_{mn}$ for the time-dependent detuning are smaller than the one for the constant detuning around $t=0$, and the peaks of $h_{mn}$ for the time-dependent detuning is lower than
the maximum value for the control with the constant detuning.

Now, a comment is in order. Using larger constant detuning also can improve the fidelity of the creation of a cat state. However, finite $\Delta$ causes $R_x$ gate of the parametron as explained in the following section because $\Delta$ increases the gap between the highest level and the second highest level of the parametron. Although using larger pump strength can decrease the gap, it increases the disturbance of the state due to NROTs. Therefore, it is favorable to make $\Delta$ zero at the end of the creation of a cat state from the point of view of the information processing.

The decay from the parametron, which decoheres the qubit state, is an another origin of the imperfection of the control.
The effect of the decay to the creation of a cat state is examined in Supplementary Section S2, although we focus mainly on the effect of NROTs in this paper.

Before moving to the next section, we surmmarize the trade-off relations and explain the role of our method.
Creation of a cat state should be followed by some other controls such as gate operations and a measurement in applications. Therefore, the speed of creation of cat state should be sufficiently faster than the decay rate for practical purposes. Moreover, for quantum computation, such a fast control is essential to improve the clock frequency.
$\beta/\chi$ should be increased rapidly, and its final value should be sufficiently large to use $| -\alpha\rangle$ and $| \alpha \rangle$ as qubit state. 
Then, we have the  trade-off relations:
1. Choosing smaller $\chi$ causes more nonadiabatic transitions due to a smaller energy gap between the eigenenergies;
2. Making $\beta$ larger causes more decrease and larger fluctuation of the fidelity of the control due to the effect of NROTs.
The modified method with the time dependent detuning can increase the fidelity by decreasing the nonadiabatic transitions and can suppress the fluctuation of the fidelity.

\subsection*{$R_x(\frac{\pi}{2})$ gate}
A pulsed detuning realizes a rotation of a parametron around the $x$ axis~\cite{Goto2016b}.
The detuning enlarges the energy difference between the highest and the second highest levels of the instantaneous Hamiltonian.
Thus, the states obtain the different dynamical phases, which give rise to a $R_x$ gate.
This scheme of the $R_x$ gate differs from the one which utilizes the time-dependent pump strength in Ref.~\citenum{Grimm2020}.

We examine the degradation of the fidelity of the $R_x(\frac{\pi}{2})$ gate due to the NROTs using the pulsed detuning given by 
\begin{eqnarray}
\Delta(t) =
\left\{ 
\begin{array}{cl}
 \Delta_0 \sin^2 ({\pi t}/{T_g})  \ & \ {\rm for}  \ \ 0\le t \le T_g,\\
0  & \  {\rm for}  \ \ t >T_g,
\end{array}
\right.
\end{eqnarray}
where $T_g$ is the gate time and $\Delta_0$ is optimized for $R_x(\frac{\pi}{2})$ gate (the angle of rotation is determined by $\Delta_0$).
The other parameters are fixed during the control.
The initial state is set to be 
\begin{eqnarray}
|\Psi(0)\rangle = (|\varphi_0\rangle + |\varphi_1\rangle)/\sqrt{2}\simeq |-\alpha\rangle.
\end{eqnarray}
The fidelity of the gate is defined by the population of the target state,
\begin{eqnarray}
|\Psi_{\rm tar}\rangle = (|\varphi_0\rangle - |\varphi_1\rangle)/\sqrt{2}\simeq |\alpha\rangle
\end{eqnarray}
at $t=T_g$.

\begin{figure}
\begin{center}
\includegraphics[width=10cm]{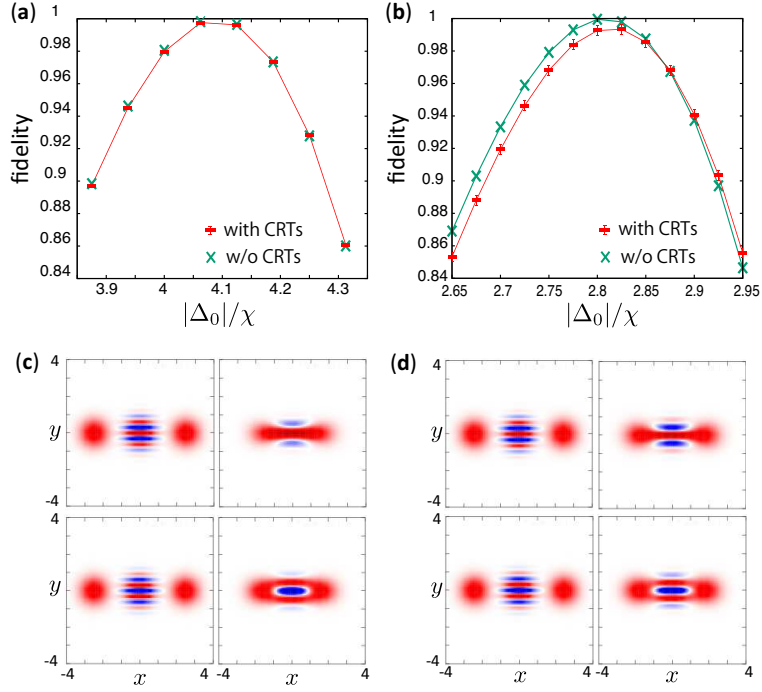}
\end{center}
\caption{
Fidelity of $R_x(\frac{\pi}{2})$ gate. The used parameters are $\beta/2\pi=53$~MHz, $\chi/2\pi=17$ MHz for panel (a) and $\beta/2\pi=200$~MHz, $\chi/2\pi=68$ MHz for panel (b).
We used $\omega_p/2\pi=16$ GHz and $T_g=100$ ns for the both panels. 
The error bars represent the standard deviation calculated using the data for $t>T_g$.
Panels (c) and (d): Wigner function of the highest (upper panels) and the second highest levels (lower panels) of $H_{\rm RWA}$ in Eq.~(\ref{H_6_8_20}).
The left and the right panels correspond to $\Delta=0$ and $\Delta=\Delta_0$, where $|\Delta_0|/\chi=4.1$ for (c) and 2.8 for (d), respectively. 
The other parameters used in panels (c) and (d) are the same as panels (a) and (b), respectively.
}
\label{pop_up2_6_8_20}
\end{figure}

We consider two sets of $(\beta,\chi)$ which give approximately the same $\alpha$.
Figures \ref{pop_up2_6_8_20}(a) and \ref{pop_up2_6_8_20}(b) show the fidelity of the $R_x(\frac{\pi}{2})$ gate for the both parameter sets with and without the NROTs.
In the case without NROTs, the both parameter sets give the fidelity of approximately unity.
The maximum fidelity for the smaller $\beta$ and $\chi$ is approximately the same as the case without the NROTs [Fig.~\ref{pop_up2_6_8_20}(a)] (The difference between them is less than 0.1\%).
On the other hand, the fidelity for the parameter set with larger $\beta$ and $\chi$ is degraded when the NROTs are taken into account as seen in Fig.~\ref{pop_up2_6_8_20}(b).
This means that, smaller parameter set is more suitable to decrease the disturbance by the NROTs in the $R_x$ gate, although the smaller parameter set tends to induce more  nonadiabatic transitions during the creation of the cat state.
Fortunately, we have found that the method with Eq.~(\ref{Delta_9_7_20}) suppresses the nonadiabatic transitions and the fluctuation of the state when we create a cat state, as shown in  Fig.~\ref{eng_cat_pop_9_3_20}.
Therefore, we can safely choose the smaller parameter set of $\beta$ and $\chi$ to achieve the higher fidelity of $R_x$ gate while the nonadiabatic transitions and the fluctuation of the state during the cat-state creation are still significantly suppressed by using the modified method.

A comment on the intermediate state during the gate operation is in order.
The larger parameter set gives small values of $|\Delta_0|/\chi$ and $|\Delta_0|/\beta$ to perform the $R_x(\frac{\pi}{2})$ gate.
The required value of $|\Delta_0|/\chi$ is approximately 4.1 and 2.8 for the smaller and the larger parameter sets, respectively. Thus, the intermediate states during the gate operations are different.
Figures~\ref{pop_up2_6_8_20}(c) and \ref{pop_up2_6_8_20}(d) show the Wigner function of the highest and the second highest levels of $H_{\rm RWA}$ in Eq.~(\ref{H_6_8_20}) for $\Delta=0$ and $\Delta=\Delta_0$.
The Wigner function, which is separated in three parts for $\Delta=0$, is connected near the origin for $\Delta=\Delta_0$.
It represents that the highest and the second highest levels become closer to the zero photon and the one photon Fock states, respectively.
The Wigner function for the larger parameter set is shrunk in the $y-$direction around the origin compared to that for the smaller parameter set because of the difference in $|\Delta_0|/\chi$.

$R_z$ and $R_x$ gates can consist of a universal single-qubit gate set. $R_z$ gates for a parametron can be realized by a drive with a microwave pulse~\cite{Goto2016b}. 
Because the intensity of the microwave pulse is sufficiently smaller than the pump field, the interplay between the microwave pulse and the NROTs is negligible (see S3 for detail). 

\section*{Conclusion}
\label{Conclusion}
We have quantitatively investigated the effect of the non-resonant rapidly oscillating terms (NROTs) on controls of a parametron.
It has been shown that the NROTs cause unwanted population transfer 
from the qubit levels to the other energy levels, and degrade the fidelity of the cat-state creation.
The population transfer is mainly from the highest level to the third highest level when the frequency of the pump field is sufficiently high.
However, we can increase the control fidelity by suitably choosing parameters 
such as the nonlinearity parameter, the pump strength and frequency.
Furthermore, starting from large detuning and decreasing it to zero as the pump is ramped, we can greatly enhance the fidelity of the cat-state creation, which we call a modified method.
Interestingly, the fluctuation of the population of the target state is suppressed in the modified method. 
The mechanism of the enhancement of the fidelity has been explained from the viewpoint of the adiabatic condition.
%\textcolor{blue}{This proposed method is useful also to avoid unwanted $R_x$ gate caused by finite detuning after the pump is ramped.}
Also, we  have studied the effect of the NROTs on a $R_x$ gate.
The fidelity of the $R_x$ gate depends on the pump strength because of the NROTs.
We  have shown that smaller pump field and nonlinearity parameter realize higher gate fidelity.

Turning on and off the pump field can be used not only for the cat-state creation but also for transforming a parametron to a transmon for the qubit readout~\cite{Grimm2020}. Therefore, the inverse process of the modified adiabatic method of the creation of a cat state is expected to be useful also for that purpose.

\section*{Acknowledgements}
This paper is partly based on results obtained from  a project,
JPNP16007, commissioned
 by the New Energy and Industrial Technology Development Organization (NEDO), Japan. 
S.M. acknowledges the support from JSPS KAKENHI (grant number 18K03486). 
Y. M. was supported by Leading Initiative for Excellent Young Researchers MEXT Japan and JST presto (Grant No. JPMJPR1919). 
The authors thank T. Yamamoto, T. Yamaji and A. Uchiyama for fruitful discussions.

\section*{Author contributions statement}
S.M. provided the initial ideas and performed numerical simulations. S.M., T.I. and Y.M. contributed to theoretical analysis. S.K. supervised the work in all respects. All authors reviewed the manuscript. 

%\section*{Competing interests}
%The authors declare no competing interests.

\clearpage
\setcounter{figure}{0}
\renewcommand*{\thefigure}{S\arabic{figure}}
\setcounter{table}{0}
\renewcommand*{\thetable}{S\arabic{table}}
\setcounter{equation}{0}

\begin{center}
{\LARGE \bf Supplemental information:\\
Controls of a superconducting quantum parametron under a strong pump field}\\ \vspace{0.4cm}

{\Large Shumpei Masuda$^{1,\ast}$, Toyofumi Ishikawa$^{1}$, Yuichiro Matsuzaki$^{1}$ and\\ Shiro Kawabata$^{1}$}\\
\vspace{0.4cm}
$^{1}$ Research Center for Emerging Computing Technologies (RCECT), National Institute of Advanced Industrial Science and Technology (AIST), 1-1-1, Umezono, Tsukuba, Ibaraki 305-8568, Japan\\

\vspace{0.4cm}
$^\ast$ shumpei.masuda@aist.go.jp
\end{center}

\section*{S1 Controls with various values of $\Delta_0$}
\label{Controls with various values of Delta 0}
\begin{figure}[h]
\begin{center}
\includegraphics[width=6.5cm]{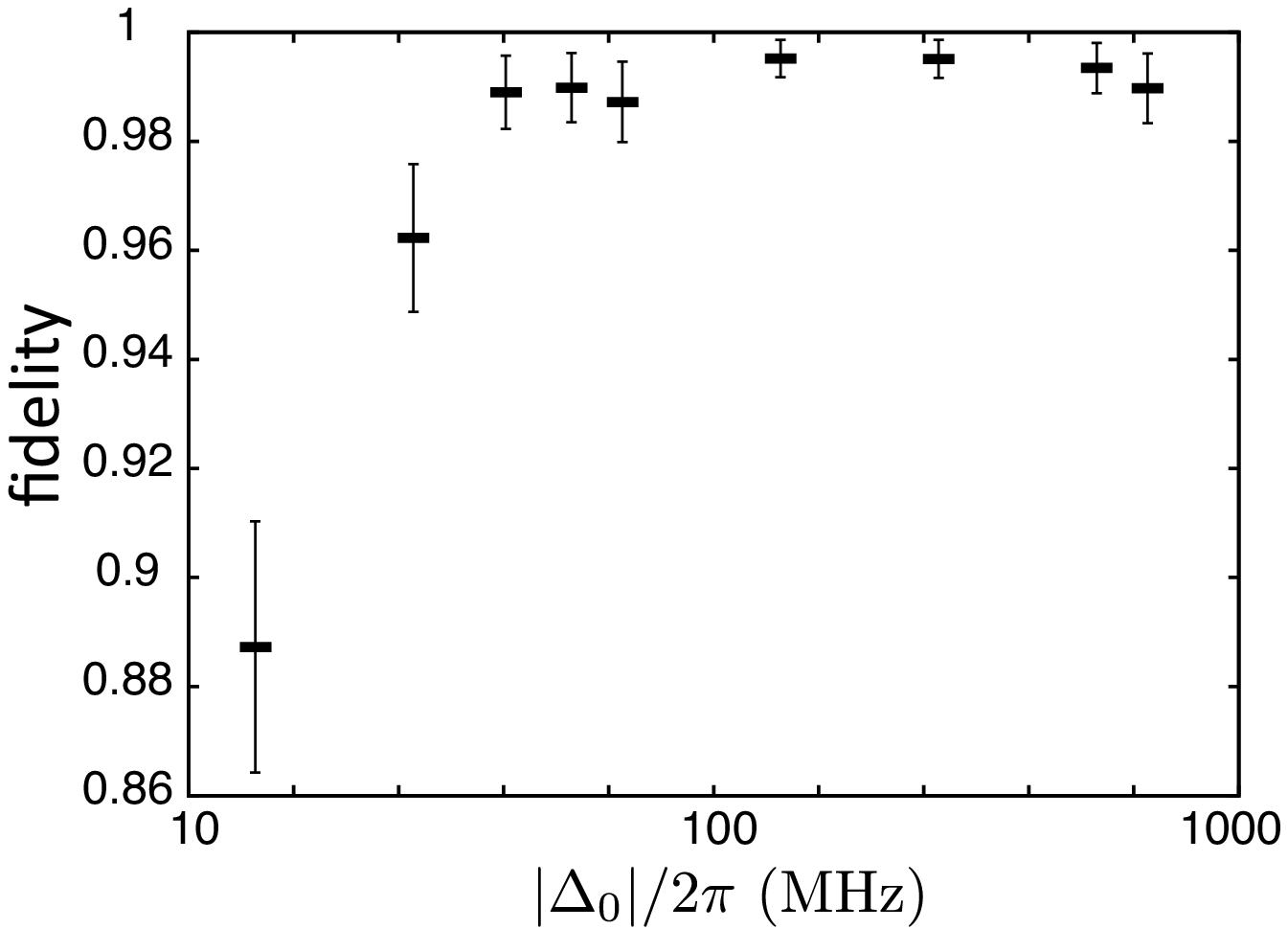}
\end{center}
\caption{
Dependence of the fidelity on $|\Delta_0|$ for the creation of a cat state for $T=20$~ns with the time-dependent detuning in Eq.~(\ref{Delta_9_7_20}) .
Other parameters are the same as Fig.~\ref{eng_cat_pop_9_3_20}(a).
}
\label{pop_data_Delta_t_12_16_20}
\end{figure}

We consider the creation of a cat state with the time-dependent detuning in Eq.~(\ref{Delta_9_7_20}) for various values of $\Delta_0$.
Figure \ref{pop_data_Delta_t_12_16_20} shows the fidelity as a function of $\Delta_0$ for $T=20$~ns.
The fidelity higher than 0.98 is realized for $|\Delta_0|/2\pi>40$~MHz.
(The fidelity for the constant detuning shown in Fig.~\ref{eng_cat_pop_9_3_20}(a) is less than 0.85.)
The fidelity does not increase monotonically with respect to $\Delta_0$ for $|\Delta_0|/2\pi>40$~MHz.
We attribute this fluctuation to nonadiabatic transitions due to the rapid change of the detuning.

\begin{figure}
\begin{center}
\includegraphics[width=10cm]{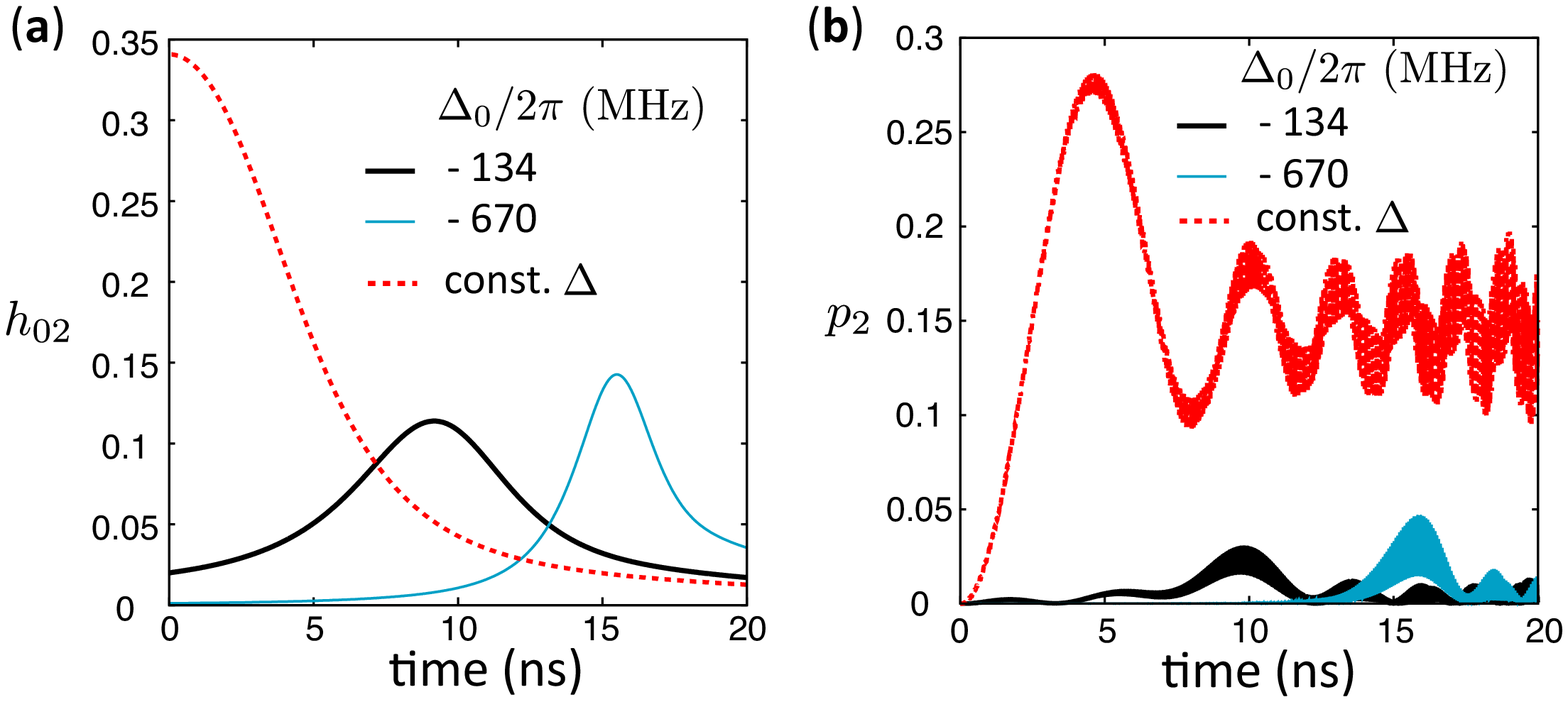}
\end{center}
\caption{
(a) Time dependence of $h_{02}$ for the creation of a cat state for $T=20$~ns with the time-dependent detuning in Eq.~(\ref{Delta_9_7_20}) and the constant detuning of $\Delta/2\pi=-6.7$~MHz.
We used $\Delta_0/2\pi=-134$~MHz and $-670$ MHz for the controls with the time-dependent detuning.
(b) Time dependence of the population of the third highest level, $p_2$, in the same dynamics as panel (a).
Other parameters are the same as Fig.~\ref{eng_cat_pop_9_3_20}(a).
}
\label{pop_level3_com_12_16_20}
\end{figure}

Figure~\ref{pop_level3_com_12_16_20}(a) shows the time dependence of $h_{02}$ in Eq.~(\ref{hmn_12_15_20}) for the control with the time-dependent detuning in Eq.~(\ref{Delta_9_7_20}) and the control with the constant detuning.
$h_{02}$ for the control with the constant detuning is high around $t=0$, and its maximum value is higher than the peaks of the controls with the time-dependent detuning. 
The peak for the time-dependent detuning with $\Delta_{0}/2\pi=-670$~MHz is slightly higher than the one for the control with $\Delta_{0}/2\pi=-134$~MHz.
The time dependence of the population of the third highest level, $p_2$, is shown in Fig.~\ref{pop_level3_com_12_16_20}(b).
It is seen that $p_2$ increases around $t=0$ in the control with the constant detuning.
On the other hand, the increase of $p_2$ occurs later in the controls with the time dependent detuning.
The average value of $p_2$ around $t=20$~ns is much smaller than the one for the control with the constant detuning.
We attribute this time dependence of $p_2$ to the time dependence of $h_{02}$ which should be small to avoid the nonadiabatic population transfer from the highest level to the third highest level.

\section*{S2 Effect of decay}
\label{Effect of decay}
In the main text, we consider controls of which duration is sufficiently shorter than the coherence time.
Thus, the decoherence of the system is neglected.
When the above condition is not satisfied, the control is degraded by the decoherence.
It is known that the decay of the the nonlinear resonator causes the effective dephasing of a parametron.
The rate of the phase decay is represented as~\cite{Puri2017bv2} $\gamma = 2\kappa |\alpha|^2$, where $\kappa$ is the decay rate of the nonlinear resonator.
In this section, we examine the effect of the decay to the two kinds of the creation of a cat state studied in the main text by solving the master equation
\begin{eqnarray}
\dot\rho = -i[H_{\rm RWA},\rho] + \frac{\kappa}{2} \Big{(} [a\rho,a^\dagger] + [a,\rho a^\dagger] \Big{)},
\end{eqnarray}
where $\rho$ is the density matrix of the system.
The NROTs are not taken into account in this calculation.

\begin{figure}
\begin{center}
\includegraphics[width=7cm]{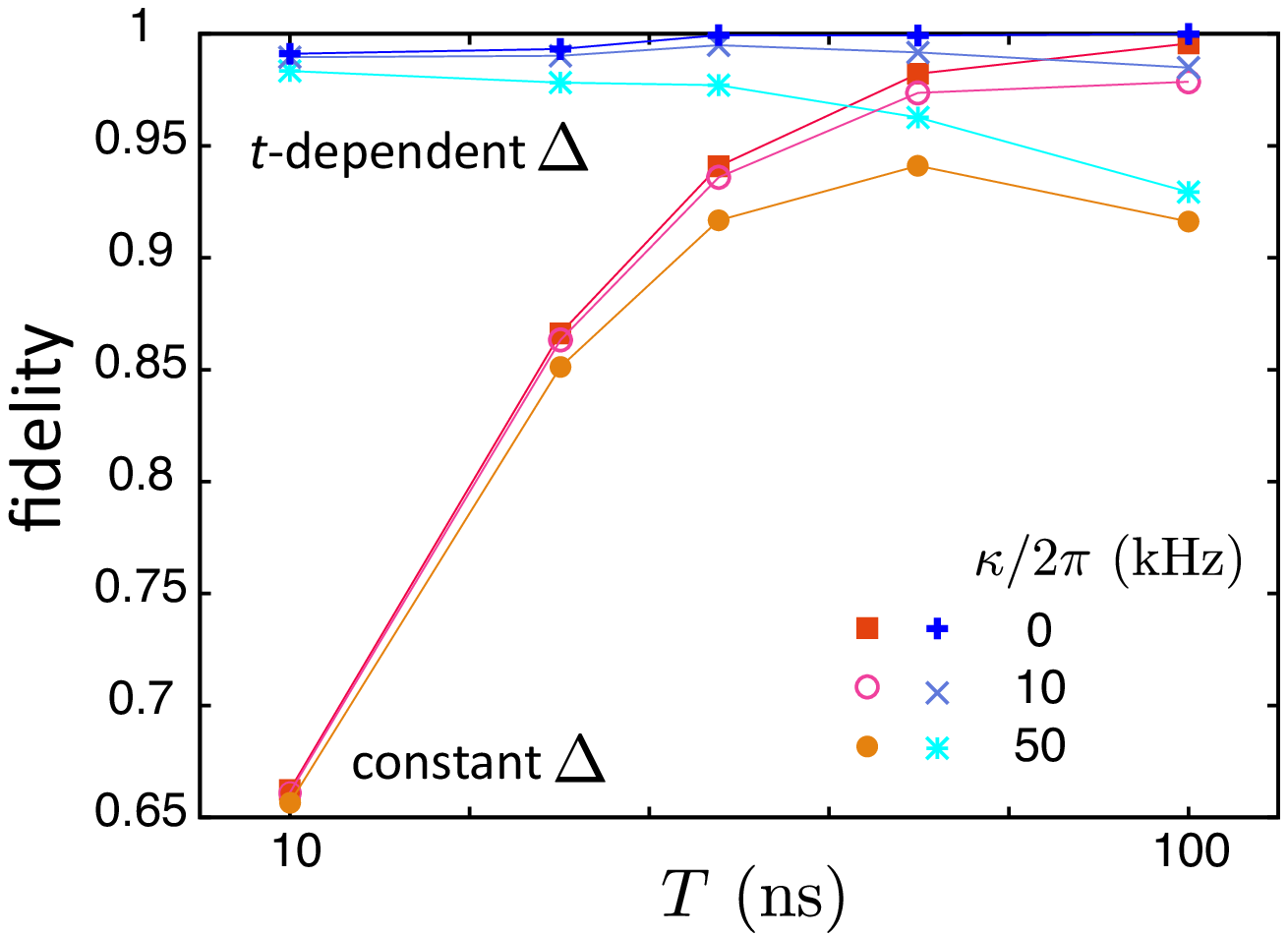}
\end{center}
\caption{
Fidelity of the creation of a cat state as a function of $T$ with the constant detuning and the time dependent detuning.
The decay rate, $\kappa$, are taken into account in the both dynamics.
The solid lines are guide to the eye.
Other parameters are the same as Fig.~\ref{eng_cat_pop_9_3_20}(a).
}
\label{fid_com_12_23_20}
\end{figure}

Figure \ref{fid_com_12_23_20} shows the fidelity as a function of $T$ for various values of $\kappa$.
It is observed that the decrease of the fidelity due to the decoherence becomes small as $T$ decreases.
For example, the change of the fidelity is less than 0.2~\% for the controls with $T=10$~ns and $\kappa/2\pi =10$~kHz.
In Ref.~\citenum{Grimm2020v2}, the amplitude damping time $T_1$ of 15.5 $\mu$s was reported for a superconducting nonlinear resonator, which corresponds to $\kappa/2\pi \simeq 10$~kHz.

\begin{figure}
\begin{center}
\includegraphics[width=9cm]{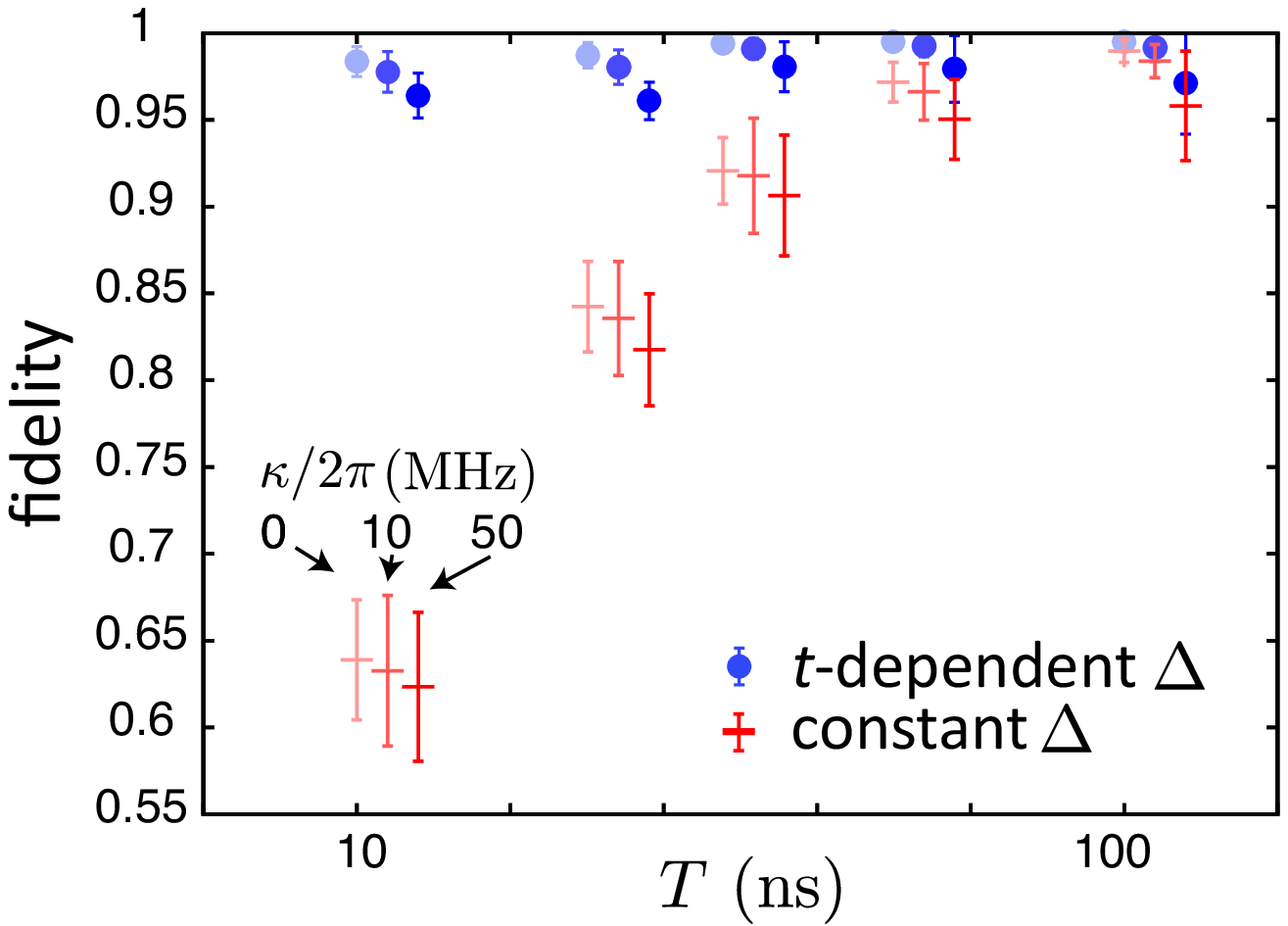}
\end{center}
\caption{
Fidelity of the creation of a cat state as a function of $T$ with the time-dependent detuning in Eq.~(\ref{Delta_9_7_20}) (blue circles) and the constant detuning of $\Delta/2\pi=-6.7$~MHz (red bars) for  $\kappa/2\pi=0$, 10 and 50~kHz. (The data points for $\kappa/2\pi=10$ and 50~kHz are shifted horizontally for clarity.)
Both the decay rate, $\kappa$, and the NROTs are taken into account in the simulation.
Other parameters are the same as Fig.~\ref{eng_cat_pop_9_3_20}(a). 
}
\label{pop_kappa_com_4_23_21}
\end{figure}

Now we take into account both the decay rate, $\kappa$, and the NROTs.
Figure~\ref{pop_kappa_com_4_23_21} shows the fidelity of the creation of a cat state for the both controls for $T=10$, 20, 30, 50, 100~ns.
The fidelity is lower compared to the cases where either of the decay or the NROTs are taken into account.
Decrease and fluctuation of the fidelity is seen even in the control with the time-dependent detuning for $\kappa/2\pi=10$~kHz and $T=10$~ns.
We attribute this to the nonadiabatic transitions because the decay becomes more significant if Fock states with higher photon number are populated. Such decease and fluctuation are relatively suppressed for $T=30$, 50 and 100 for $\kappa/2\pi=10$~kHz in the control with time dependent detuning.

\section*{S3 $R_z$ gate}
\label{$R_z$ gate}
$R_z$ gate for a parametron using a pulsed microwave was proposed in Ref.~\citenum{Goto2016bv2}. 
Drive of a parametron by a microwave with the frequency of $\omega_p/2$ and the amplitude of $E(t)$ adds a term:
\begin{eqnarray}
H_z(t) = \hbar E(t)(a +a^\dagger )
\end{eqnarray}
into Hamiltonian (\ref{Hrf_9_1_20}) in the rotating frame used in the main text. 
When $|E(t)|$ is sufficiently small,  the parametron is approximately kept in the subspace expanded by $|-\alpha\rangle$ and $|\alpha\rangle$, where we assume that $\beta$ is constant,  $\Delta=0$, and 
$\beta/\chi$ is sufficiently large so that the overlap, $\langle -\alpha |\alpha \rangle$, is negligible. 
The energy of $|\pm \alpha\rangle$ shifts by $\pm 2\hbar E(t)\sqrt{2\beta/\chi}$.
The energy shifts give different dynamical phase to $|-\alpha\rangle$ and  $|\alpha\rangle$.
Thus, the phase difference between the two coherent states results in $R_z(\phi)$ gate at $t=T$ with
\begin{eqnarray}
\phi = 4\sqrt{2\beta/\chi}\int_0^{T} E(t) dt.
\end{eqnarray}
%In this method $|E(t)|$ is chosen to be sufficiently small so that the state of the parametron is kept in the subspace.
The interplay between the pulse for $R_z$ gate and the NROTs of the pump field can be neglected because $|E(t)|$ is much smaller than $\beta$ as shown below.

\begin{figure}
\begin{center}
\includegraphics[width=8cm]{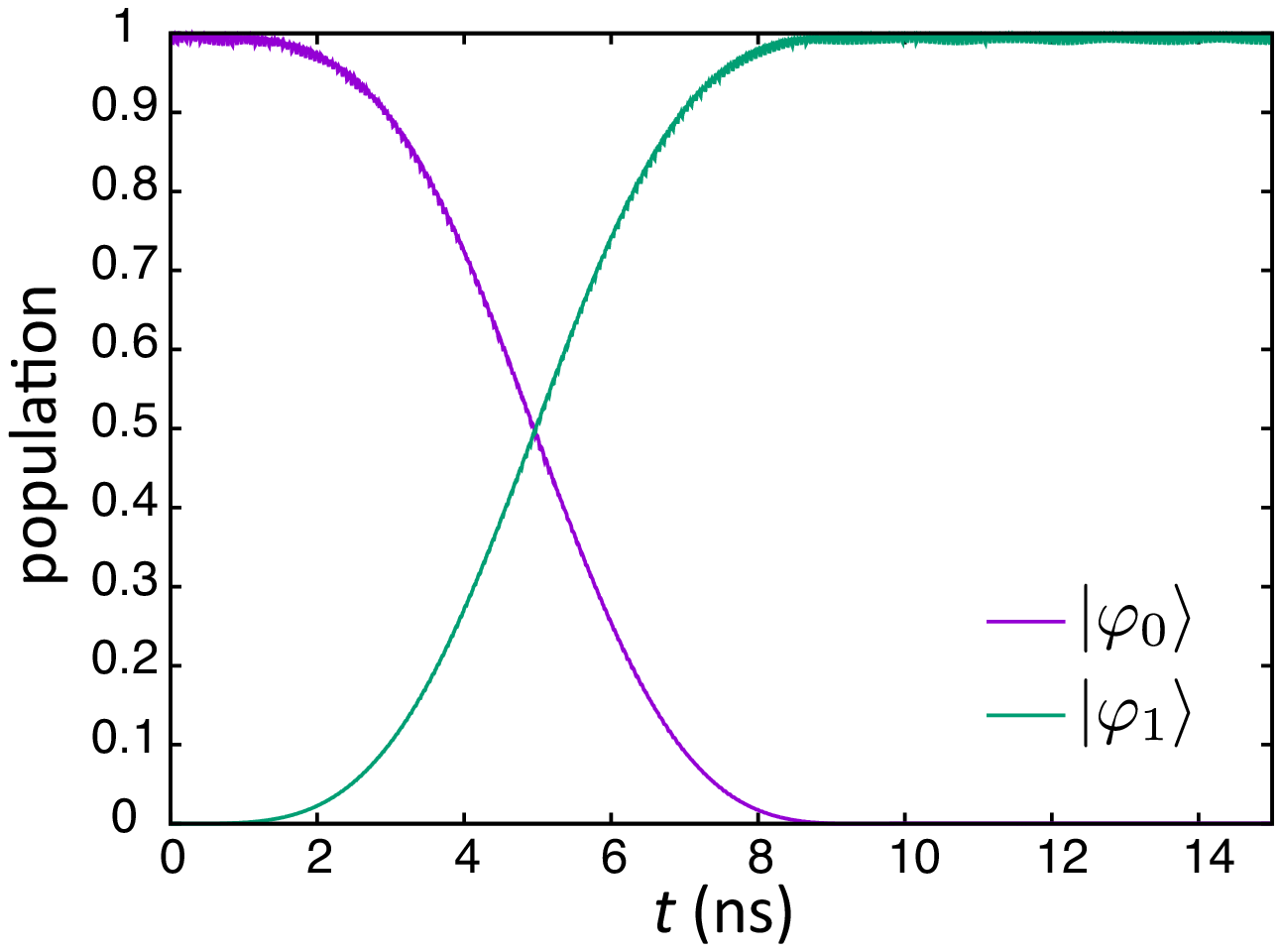}
\end{center}
\caption{
Time dependence of the population of the highest and the second highest levels during the $R_z(\pi)$ gate with $E(t)$ in  Eq.~(\ref{E_4_21_21}).
The pump strength is kept constant, $\beta(t)=\beta_0$.
The used parameter set is $T_g=10$~ns,  $\Delta/2\pi=0$~MHz, $\beta_0/2\pi = 200$~MHz, $\omega_p/2\pi=16$~GHz, $\chi/2\pi=68$~MHz.
} 
\label{pop_com_4_21_21}
\end{figure}

Now, we simulate $R_z(\pi)$ taking into account NROTs of the pump field using $E(t)$ given by \cite{Goto2016bv2}
\begin{eqnarray}
E(t) =
\left\{ 
\begin{array}{cl}
\frac{\pi^2}{8T_g \sqrt{2\beta/\chi}} \sin \frac{\pi t}{T_g}  \ & \ {\rm for}  \ \ 0\le t \le T_g,\\
0  & \  {\rm for}  \ \ t >T_g,
\end{array}
\right.
\label{E_4_21_21}
\end{eqnarray}
where $T_g$ is the duration of the pulsed field.
The pump field is fixed as $\beta=\beta_0$.
The initial state is the highest level, $|\varphi_0\rangle$, in Eq.~(\ref{cat_6_23_20}).
We use the parameter set: $T_g=10$~ns,  $\Delta/2\pi=0$~MHz, $\beta_0/2\pi = 200$~MHz, $\omega_p/2\pi=16$~GHz, $\chi/2\pi=68$~MHz. The peak value of $|E(t)|$ is 24 times smaller than $\beta_0$.
The $R_z(\pi)$ gate drives the state to the second highest level, $|\varphi_1\rangle$, in Eq.~(\ref{cat_6_23_20}). The fidelity of the control is defined by the population of $|\varphi_1\rangle$. The fidelity averaged for $t>T_g$ and the standard deviation of the fluctuation of the fidelity are 0.994 and 0.003, respectively.
Figure~\ref{pop_com_4_21_21} shows the time dependence of the population of the highest and the second highest levels for $0<t<15$~ns. It is seen that the population is transferred from $|\varphi_0\rangle$ to $|\varphi_1\rangle$. 

We simulate the dynamics without the pulsed field, $E(t)=0$, for comparison. In this dynamics, the parametron should stay in $|\varphi_0\rangle$ if there is no NROT. The fidelity is defined by the population of $|\varphi_0\rangle$. The averaged fidelity and the standard deviation of the fluctuation of the fidelity are approximately the same as those with the pulsed field in Eq.~(\ref{E_4_21_21}). Therefore, the effect of the interplay between the pulsed filed for $R_z(\pi)$ and the NROTs of the pump field can be neglected with the parameters used.

\end{document}